\documentclass[conference]{IEEEtran}
\IEEEoverridecommandlockouts
\usepackage{amsmath,amssymb,amsfonts}
\usepackage{algorithmic}
\usepackage{graphicx}
\usepackage{textcomp}
\usepackage{xcolor}
\usepackage{etoolbox}
\usepackage{cite}
\usepackage{graphicx}
\usepackage{float}
\usepackage{soul}%

\def\BibTeX{{\rm B\kern-.05em{\sc i\kern-.025em b}\kern-.08em
    T\kern-.1667em\lower.7ex\hbox{E}\kern-.125emX}}
\begin{document}

\title{\textbf{The Potential of LLMs in Automating Software Testing: From Generation to Reporting}  \\
}
\author{\IEEEauthorblockN{1\textsuperscript{st} Betim Sherifi}
\IEEEauthorblockA{
\textit{Electrical Eng. and Computer Science} \\
\textit{Florida Institute of Technology}\\
Melbourne, Florida \\
bsherifi2023@my.fit.edu}
\and
\IEEEauthorblockN{2\textsuperscript{nd} Khaled Slhoub}
\IEEEauthorblockA{
\textit{Electrical Eng. and Computer Science} \\
\textit{Florida Institute of Technology}\\
Melbourne, Florida \\
kslhoub@fit.edu}
\and
\IEEEauthorblockN{2\textsuperscript{nd} Fitzroy Nembhard}
\IEEEauthorblockA{
\textit{Electrical Eng. and Computer Science} \\
\textit{Florida Institute of Technology}\\
Melbourne, Florida \\
fnembhard@fit.edu}
}

\maketitle

\begin{abstract}
Having a high quality software is essential in software engineering, which requires robust validation and verification processes during testing activities. Manual testing, while effective, can be time consuming and costly, leading to an increased demand for automated methods. Recent advancements in Large Language Models (LLMs) have significantly influenced software engineering, particularly in areas like requirements analysis, test automation, and debugging. This paper explores an agent-oriented approach to automated software testing, using LLMs to reduce human intervention and enhance testing efficiency. The proposed framework integrates LLMs to generate unit tests, visualize call graphs, and automate test execution and reporting. Evaluations across multiple applications in Python and Java demonstrate the system's high test coverage and efficient operation. This research underscores the potential of LLM-powered agents to streamline software testing workflows while addressing challenges in scalability and accuracy.
\end{abstract}


\section{Introduction}
Similarly to other industries, ensuring that software products perform as intended and meet quality standards is a critical aspect of software engineering. Software engineers must ensure that the development artifacts are defect-free and that the software system satisfies the specified business requirements. This challenge is resolved during the software testing phase, where a cross-examination of the implementation results against the desired user requirements occurs; software testing, through validation and verification, portrays an accurate image of the quality of the software. Although ensuring a total absence of errors and defects is virtually impossible, the goal is to identify and address as many detectable errors as possible prior to the release of the product to end users \cite{industrial_surveys_on_software_testing}, \cite{software_testing_techniques}. 

Since the early 1970s, the software industry's focus on testing has grown consistently, accompanied by ongoing advancements in the methods and practices employed in the field. In fact, it is also considered one of the most important areas in the education of software engineering \cite{beyond_tech_skills_software_testing}. Due to the time-consuming and costly nature of manual testing, particularly for regression testing, the need for effective automated testing approaches has increasingly emerged. However, both manual and automated approaches have the potential to improve or restrict the efficiency of the test. Although automated testing may appear to be a straightforward alternative to manual methods, it is not a complete substitute. For example, in GUI testing, automation tools can help with specific tasks, but cannot provide a fully comprehensive testing process without human involvement \cite{trade_offs}.

A software agent is a flexible and autonomous entity that interacts directly with their environment, receives input from sensors, and influences it through actuators \cite{intro_multi_agent_systems}. A multi-agent system (MAS), in contrast, comprises a collection of distinct agents that collaborate through communication and information exchange to accomplish tasks that exceed the capabilities of a single agent.

Significant advancements in machine learning and natural language processing techniques, combined with increased computational power and the availability of extensive training datasets, have driven the development of Large Language Models (LLMs). These models are capable of generating human-like text, often making the distinction between machine-generated and human-created content challenging. The advent of LLMs has proven to be highly beneficial across various domains of software engineering, including software implementation (e.g., code completion, generation, and summarization), software maintenance, quality assurance, and requirements engineering. For example, in requirements engineering, tools such as ChatGPT and Gemini are particularly effective in resolving ambiguities in specifications and categorizing requirements as either functional or non-functional. \cite{hou2024largelanguagemodelssoftware}. 

Building on the capabilities of LLMs and using recent advances in automated testing tools, this research proposes a multi-agent-based software testing approach powered by advanced LLMs. Unlike traditional automated methods where test cases are written by software developers or testers, our approach leverages LLMs to generate tests dynamically, significantly reducing the need for manual input and minimizing the time required for test case creation and execution. The primary objective is to reduce reliance on human intervention while delivering a comprehensive framework for automated software testing. The research aims to demonstrate how integrating LLMs into a multi-agent system can transform software testing from a primarily reactive and manual process to a proactive, intelligent, and more autonomous quality assurance mechanism. This approach seeks to address critical challenges in software testing, including test case design and comprehensive coverage, by generating test cases, identifying edge scenarios, and providing a rationale behind the tests.
 
The structure of this paper is organized as follows. Section 2 provides a discussion of related prior work. Section 3 outlines the proposed methodology and summarizes the results. Section 4 evaluates the proposed methodology through case studies. Finally, Section 5 concludes the paper with final remarks.

\section{Related Work}

This section presents a review of pertinent research studies in the domain of software testing, with a particular focus on Agent-Based Software Testing (ABST). The analysis encompasses existing methodologies, including those LLM Models as well as more conventional approaches that do not incorporate such models.

\subsection{Traditional Approaches to Agent-Based Software Testing }

Agent-Based Software Testing (ABST) refers to the application of agent-based systems—encompassing software agents, intelligent agents, autonomous agents, and multi-agent systems—to address challenges in software testing. ABST aims to enhance the testing process by automating complex and resource-intensive testing tasks, thereby improving efficiency and accuracy \cite{ABST_systematic_mapping_study}. This approach has attracted interest since 1999, with research activity reaching notable peaks over the past decade. The majority of ABST studies emphasize system-level testing, with Java serving as the predominant target language. Nevertheless, other programming languages, such as C, C++, Python, and Perl, have also been explored. In \cite{ABST_Web_Based_Systems}, the authors propose an agent-based framework designed for the automated testing of web-based systems. The framework incorporates multiple agents, including the Test Runtime Environment (TRE) agent, which serves as the central component of the system, as well as the Test Script Generator, Test Executor (TE) agents, and the Dashboard agent. These agents work collaboratively to execute and monitor the testing processes. Similarly, \cite{multi_agent_web_based} introduces a multi-agent-based software environment tailored for web-based applications, in which agents can dynamically join or leave the system. Communication among agents is structured across three layers: the lowest layer facilitates message transmission between agents, the middle layer defines message content and ontology, and the highest layer employs communication protocols grounded in speech-act theory. An alternative approach is presented in \cite{Software_Agents_Prioritization}, where the authors developed multiple agents to test an industrial coffee machine. Each agent represents a software module and employs fuzzy logic to prioritize software tests based on fault probabilities.  

\subsection{Enhancing Agent-Based Software Testing with Large Language Models (LLMs)}

A study by \cite{Survey_LLM_Testing} explores the practical application of Large Language Models in software testing within the software industry. Drawing insights from 83 completed questionnaires submitted by experienced and diverse professionals in the field, the research seeks to understand the extent and manner in which software testing practitioners utilize LLMs across different phases of the software testing lifecycle. The anonymous survey gathered data on participants' usage of LLMs in testing activities, the specific tools employed, and their perspectives on the potential of LLMs in this domain. The findings indicate that while 52\% of respondents have not yet utilized LLMs for testing, 48\% have integrated these models into various activities, including requirements analysis, test plan development, and test automation. Tools such as ChatGPT and GitHub Copilot were frequently mentioned. The study concludes that LLMs hold significant promise as tools for enhancing software testing practices. However, the authors caution against indiscriminate usage, particularly in scenarios involving sensitive information, underscoring the need for careful consideration in such contexts. A recent comprehensive review by \cite{software_testing_llm_survery_landscape} analyzed 102 research papers, including 82 published in 2023, focused on the application of Large Language Models (LLMs) in software testing. The review highlights the utility of LLMs in mid-phase software testing activities, such as generating unit test cases, crafting test assertions, and producing system test inputs. Furthermore, LLMs are identified as valuable tools in later stages of the software testing lifecycle, providing support for bug analysis, debugging, and automated repair processes. In \cite{towards_llm_assisted_system_testing}, the researchers introduced Kashef, a tool designed to assist software testers in designing, generating, and executing test cases, with a particular focus on web microservices applications. Kashef employs a multi-agent architecture comprising two agents powered by Large Language Models (LLMs)—the Test Engineer and the HTML Interpreter—alongside a non-LLM agent, the Code Executor. The tool integrates various LLMs, including GPT-3.5, GPT-4, CodeLlama, and Llama2, in conjunction with supporting libraries such as LangGraph for machine learning functionalities and Selenium for testing purposes.

\section{Proposed Methodology}

\subsection{High Level Architecture Description}
\begin{figure*}[htbp]
\centerline{\includegraphics[width=0.8\linewidth]{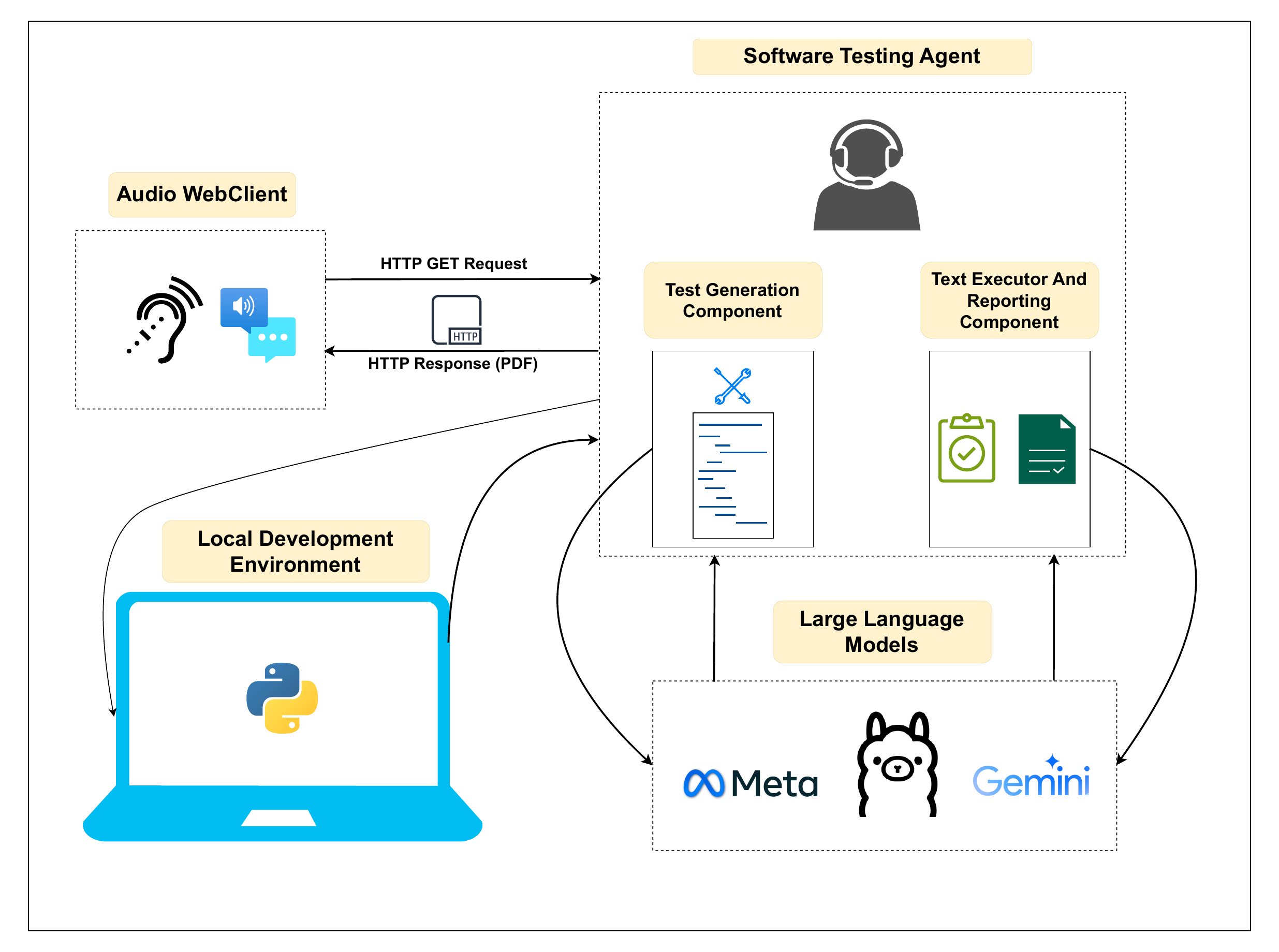}}
\caption{The high-level architecture of the proposed framework}
\label{fig_high_level_arch}
\end{figure*}

As depicted in Fig. \ref{fig_high_level_arch}, the proposed architecture consists of four primary components, each playing a unique role in coordinating and enhancing the automated software testing framework.

\begin{enumerate}
    \item \textbf{Audio WebClient}:
    A foundational web application designed to capture user inputs, either through voice commands or text. Its primary role is to initiate the testing workflow by sending an HTTP GET request to the Software Testing Agent.
    \item \textbf{Software Testing Agent}:
    It serves as the system's core component, hosting the primary logic and coordinating interactions between various components. The Software Testing Agent functions as an abstraction layer for smaller subcomponents tasked with key operations such as generating test scripts, executing tests, and creating bug reports.
    \item \textbf{Large Language Models (LLMs)}:
    LLMs are integral to the proposed automated testing framework. The Software Testing Agent leverages LLMs for various tasks, including extracting key entities from user commands to streamline operations, generating customized unit tests tailored to specific requirements, and producing DOT graphs to visualize the application's call graph.
    \item \textbf{The Development Environment}:
    It serves as the workspace housing the project code to be tested. This component facilitates automated testing by providing access to the target project, executing generated test cases, and displaying testing results and coverage metrics.
\end{enumerate}

\subsection{Low Level Architecture Description}

\begin{figure*}[htbp]
\centerline{\includegraphics[width=0.9\linewidth]{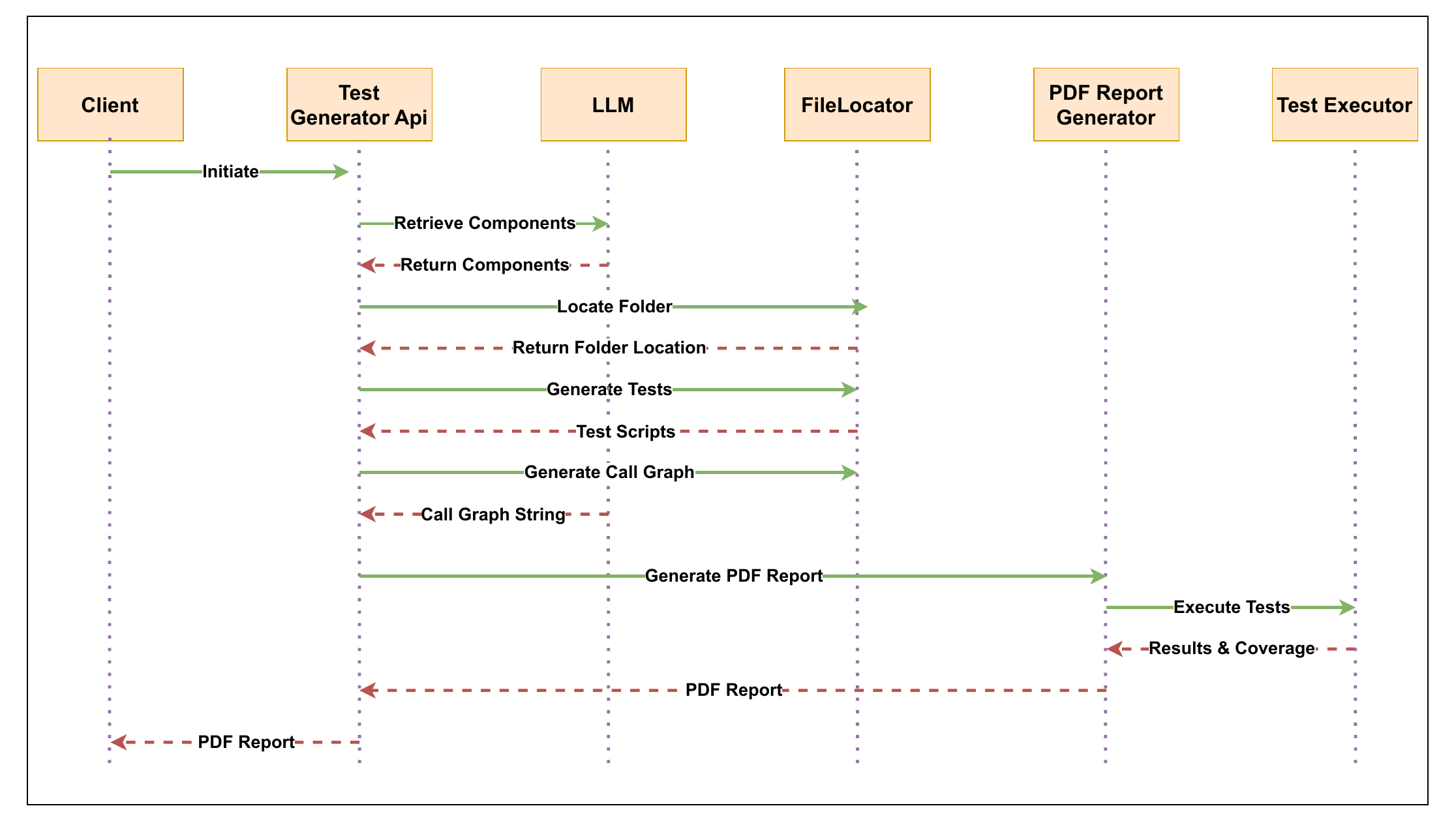}}
\caption{The low-level architecture of the proposed framework}
\label{fig_low_level_arch}
\end{figure*}

The low-level architecture, illustrated in Fig. \ref{fig_low_level_arch}, provides a detailed overview of the entire workflow, highlighting the responsibilities of each module and their interactions within the system. The process is initiated by the \textbf{Client}, typically through a voice command. The user prompt is received by the \textbf{Test Generator API}, which represents the entry point for the testing framework. The prompt is forwarded to the integrated \textbf{LLMs}, where entities such as the project name, the specific subfolder and the programming language of the project are extracted. Once the LLM identifies those entities, the \textbf{FileLocator} module is activated, which locates the specified project folder and locates the files within it.

Following that, the contents of these files are extracted and are sent as a prompt to the LLM, with the objective of generating unit tests and their accompanying rationale, using Gemini in our case. Additionally, the LLM generates a DOT graph string to help visualize a call graph of the code base interactions. After receiving the generated test scripts, the test files are created within the project. Next, the \textbf{PDF Report Generator} is activated, which as a first step orders the \textbf{Test Executor} to execute the newly created test files. Details results of the test runs, including code coverage metrics are provided by the Test Executor. Upon receiving this information, the Report Generator creates a comprehensive PDF report including test results, the rationales, code coverage and the generated call graph. This report is then returned to the user, finalizing the whole workflow.

\section{Case Study and Results}

To evaluate the performance of this system, numerous experiments were executed across four different applications. Two of the projects were implemented in Python and the other two in Java. The smaller applications included a Python application, labeled Experiment, offering basic calculator functionalities such as addition, multiplication, and Fibonacci sequence generation, and a Java application, StudentAverage, designed to calculate students' average grades across courses. The more complex projects were also split by language: a Python-based cinema management system facilitating movie rentals and a Java-based library management system with analogous features. Prompts of various formats were used to initiate the application’s execution. Examples include requests such as \textit{“Please create unit tests for the project Library under the folder management, written in Java"}, and \textit{"Write Python-based tests for the cinema project, specifically for the models folder.”} 

\subsection{Report description}

The generated output of the system is a structured PDF report, presenting detailed insights regarding the testing process and results. It begins with listing the \textbf{Test Rationale} of each file and its functions. For every function, it describes the basic test cases and any applicable edge cases. This is followed by the \textbf{Test Results} section, which lists all executed test cases. Each test case is marked as Passed (highlighted in green) or Failed (highlighted in red), with additional details provided for any failures. Next, the \textbf{Coverage Table} summarizes the testing coverage, shoeing percentages, statements tested, and missed statements for each test file, along with an overall summary. Finally, the report concludes with a Call Graph that visualizes the code interactions. For future improvements, the call graph could be enhanced with a heatmap overlay, highlighting areas more prone to errors and requiring additional testing.

For example, in the cinema management project, the rationale for the "\textbf{test rent movie}" function in the Library class includes the following scenarios:

\begin{itemize}
    \item Basic case: \textit{“Tests renting an available movie to an existing member.”}
    \item Edge cases: \textit{“Tests renting a non-existent movie, renting a movie to a non-existent member, and renting an already rented movie.”}
\end{itemize}
In contrast, a simpler example can be seen in the library management project for the getTitle function in the Book class:

\begin{itemize}
    \item Basic case: \textit{“Tests retrieving the book title after object creation.”}
    \item Edge cases: \textit{Not applicable.}
\end{itemize}

\subsection{Performance Analysis}

The system's performance was assessed by logging detailed timing information for each execution, including the duration of operations such as component retrieval, unit test generation, and test execution. Additionally, test coverage and the overall execution status were recorded and categorized as either successful—when all steps, including test generation, execution, and PDF report generation, were completed successfully—or failed. Out of 20 executions on the Python projects, the system completed all runs without any failures. In contrast, 3 failures were observed out of 24 executions for the Java projects. These failures were primarily caused by ambiguous prompts, which hindered the accurate identification of necessary components, or by generated test scripts containing compilation errors, leading to failures during test execution.

The average total execution time for each run was approximately 83.5 seconds. The largest portion of this time, about 62.8 seconds on average, was dedicated to test generation by the LLM, which included generating rationales. This was followed by folder location, taking an average of 9.7 seconds, DOT graph generation at 5.4 seconds, and test execution at 3.2 seconds. Retrieving components required 1.3 seconds on average, while both PDF report creation and writing test files took less than a second each.

\begin{figure*}[htbp!]
\centerline{\includegraphics[width=0.7\linewidth]{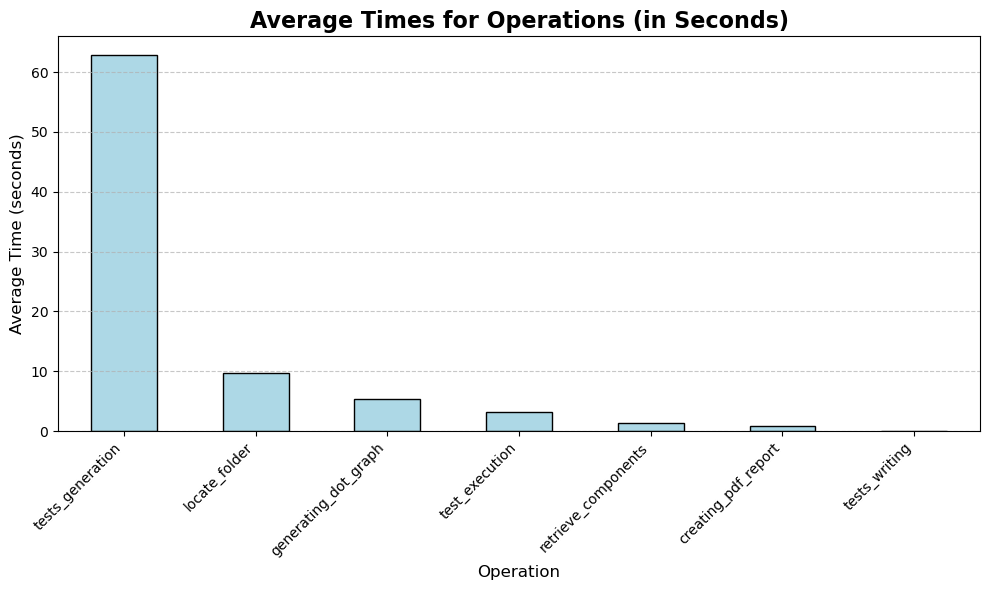}}
\caption{Average time for operations (in seconds)}
\label{operations_average_time}
\end{figure*}

\begin{table*}[htbp!]
    \centering
    \caption{Project Metrics Comparison}
    \begin{tabular}{llllcccccc}
        \hline
        \textbf{Project} & \textbf{Language} & \textbf{LoC} & \textbf{Total Time} & \textbf{Tests Gen.} & \textbf{Dot Graph} & \textbf{Entity Retrieval} & \textbf{Test Exec.} & \textbf{PDF Report} & \textbf{Coverage \%} \\
        \hline
        LibrarySystem & Java & 269 & 119.06 & 92.54 & 7.57 & 1.44 & 5.39 & 2.01 & 94.67 \\
        StudentManager & Java & 114 & 62.55 & 39.79 & 5.08 & 1.36 & 5.48 & 0.75 & 100.00 \\
        cinema & Python & 183 & 110.13  & 92.43 & 5.83 & 1.33 & 0.79 & 0.65 & 88.30 \\
        experiment & Python & 47 & 49.78 & 34.17 & 3.44 & 1.33 & 0.96 & 0.28 & 98.60 \\
        \hline
    \end{tabular}
    \label{tab:project_metrics}
\end{table*}

When comparing the average total execution time by programming language, Java executions averaged 86.7 seconds, while Python executions averaged 80 seconds. Interestingly, the time taken for test generation by the LLM was nearly identical for both languages, with Java averaging 62.4 seconds and Python 63.3 seconds. Most operations across the two programming languages showed similar average durations. However, a notable difference was observed in the test execution phase, where Java required an average of 5.44 seconds, compared to only 0.87 seconds for Python. This discrepancy in test execution time accounts for the gap in the overall average execution times between the two languages, as other operation times remained consistent, as shown in Fig. \ref{operations_average_time}.

As anticipated, the average time for test generation varied significantly across projects, with more complex projects requiring substantially more time. For example, LibrarySystem averaged 92.54 seconds, Cinema 92.43 seconds, StudentManager 39.79 seconds, and Experiment 34.17 seconds. This variation is due to the greater number of files and increased logical complexity in the more complex projects. A summary of this analysis is presented in TABLE \ref{tab:project_metrics}.

However, test execution times were relatively consistent across projects of the same programming language, even though varying in complexity. As noted earlier, Java projects had higher execution times, with LibrarySystem averaging 5.39 seconds and StudentManager 5.48 seconds. In contrast, Python projects showed much lower execution times, with Cinema averaging 0.79 seconds and Experiment 0.96 seconds.

The overall test coverage across all projects was notably high, reflecting the effectiveness of the generated test cases. When grouped by project, the coverage percentages revealed some variation. StudentManager achieved full coverage at 100\%, while Experiment followed closely with 98.6\%. LibrarySystem exhibited slightly lower coverage at 94.67\%, and Cinema recorded the lowest coverage at 88.3\%. This is summarized in TABLE \ref{tab:project_metrics}.

When analyzing the results by programming language, Java projects demonstrated slightly higher average coverage at 97.71\% compared to Python projects, which averaged 93.45\%. This difference highlights potential language-specific factors influencing coverage, such as testing frameworks or execution times. Despite this variation, both languages consistently delivered high coverage, ensuring reliable testing across all projects.

To evaluate the performance of an alternative LLM, such as ChatGPT, we tasked it with generating test cases for the LibrarySystem project. The total generation time was approximately 3 minutes, which is twice as long as the average time observed with Gemini. Despite the longer duration, the test coverage achieved was impressively high at 98\%. Notably, the ChatGPT Web Application was used for this task instead of the API, and the additional rendering process may have contributed to the increased generation time.

\section{Conclusion}

This article highlights the significant potential of LLM-powered agents in automating various software testing activities, such as test generation, execution, and reporting. The proposed framework achieved impressive success rates, with no failures observed in Python applications and a high success rate recorded for Java applications. Average execution times were 80 seconds for Python projects and 86.7 seconds for Java projects. The system also demonstrated strong test coverage, averaging 97.71\% for Java projects and 93.45\% for Python projects, showcasing its effectiveness in identifying comprehensive testing scenarios.

Despite its successes, the framework has notable areas for enhancement. Failures in Java executions, attributed to ambiguous prompts and compilation errors in generated scripts, point to the need for enhanced natural language processing and improved syntax accuracy. Moreover, the framework currently focuses exclusively on unit testing, leaving integration and system-level testing unexplored. While the call graph visualization is helpful, it could be enhanced with a heatmap overlay to identify areas more prone to defects. Incorporating requirement specification documents as prompts could further refine the precision and relevance of the generated test cases. Future work aims to address these limitations by extending language support and introducing additional test types. Despite these challenges, this research lays a solid foundation for LLM-assisted automated testing, demonstrating the potential of LLMs to minimize human intervention, boost efficiency, and improve testing quality.





\end{document}